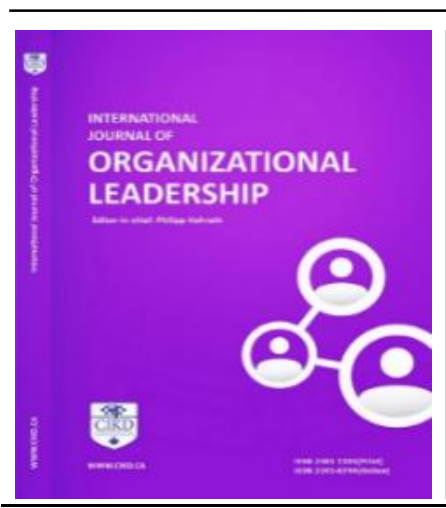

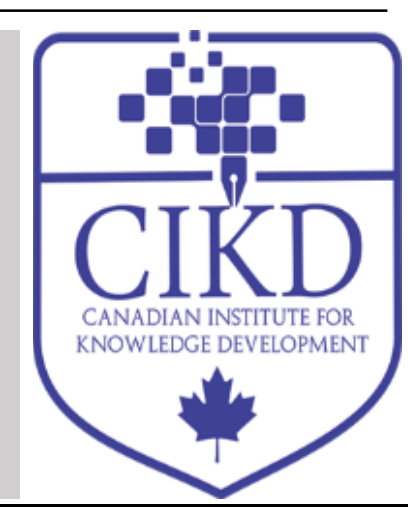

# Algorithmic Authority and the Complexities of Delegated Decision-Making: Case Studies on Ethical Challenges for 21st-Century Leadership

**Victor Frimpong**

Management Department, SBS Swiss Business School, Zurich, Switzerland

***Correspondence:**
*v.frimpong@research.sbs.edu*

## ABSTRACT

The rapid integration of AI into high-stakes decision-making has outpaced traditional mechanisms for human oversight and accountability, leaving leaders without clear guidance on how to leverage algorithmic systems responsibly. To address this gap, we conducted a comparative qualitative study of four landmark AI deployments: the UK A-Level grading algorithm used during the COVID-19 pandemic, Amazon's automated hiring tool, the COMPAS recidivism risk score in the U.S. criminal justice system, and the Dutch SyRI welfare-fraud detection system. Drawing on 61 publicly available government reports, internal memos, and media articles, we applied a rigorous two-phase grounded-theory coding process in NVivo, producing a comprehensive 32-item codebook and achieving substantial inter-coder reliability. We then quantified thematic occurrences across 110 coded segments and conducted chi-square tests to confirm consistent application of themes across cases. Our analysis yielded four actionable principles: 1) Intentionality—leaders must consciously elect to involve AI rather than default to automation; 2) Interpretability—systems should provide accessible explanations for bias detection and decision justification; 3) Moral Authorship—human actors must explicitly claim ultimate responsibility for outcomes; 4) Justice—delegation structures must be designed to prevent the perpetuation of existing inequities. Together, these principles form a reproducible analytical roadmap and offer practical guidance for accountable AI governance in high-stakes contexts.

Classical leadership theories—transformational (Bass, 1985), servant (Greenleaf, 1977), and ethical leadership (Brown & Treviño, 2006)—view human leaders as the primary moral agents in organizations. However, these theories fail to address the complexities of delegated decision-making in the AI era, where judgment is often shared with algorithmic systems (Floridi & Sanders, 2004; Subrahmanyam, 2025). Current research on leadership mainly emphasizes digital transformation and efficiency, neglecting the ethical dilemmas and accountability challenges that arise when moral responsibility is shared between humans and machines. Leaders must learn to work alongside intelligent systems. The rise of AI has changed workflows and altered leadership and decision-making processes. Algorithms now play a significant role in decisions across various fields, including education, criminal justice, healthcare, and corporate governance. AI can boost efficiency and accuracy, but it poses important ethical issues about leadership authority and accountability when machines impact decisions (Frimpong, 2025). This change goes beyond technology; it fundamentally challenges traditional leadership models.

The gap in ethical examination of algorithmic decision-making is clear in high-profile cases like the UK's A-Level grading scandal, Amazon's biased résumé screening, and the Dutch welfare fraud algorithm crisis. These situations highlight that algorithms can cause moral failures and public backlash, eroding leadership legitimacy without ethical scrutiny. Although AI ethics literature is expanding (e.g., Elish, 2019; McGuire & De Cremer, 2022; Tigard, 2020), leadership studies have not adequately addressed the consequences of algorithmic authority on moral accountability and decision-making. Interest in digital leadership is increasing (Chandra, 2025; Pavitra et al., 2024), yet we still lack clarity on how AI influences leadership authority, moral responsibility, and ethical judgment.

Leaders need a new approach to address the ethical challenges posed by AI in decision-making. This paper introduces "ethical delegation," a framework for assigning decision-making power to AI systems while maintaining moral responsibility, transparency, and human oversight. By analyzing five public case studies, the study highlights how ethical failures arise when leaders rely too heavily on opaque technologies and suggests a principled method to regain ethical control in AI-driven situations.

This study aims to investigate how leadership ethics should adapt to the growing use of AI in high-stakes decision-making. It explores how leaders can ethically delegate decision-making to algorithms while maintaining moral accountability. The key questions addressed are: How can leaders responsibly assign judgment to AI systems? What core principles define 'ethical delegation,' and how do they maintain accountability in AI-mediated decision-making? The research utilizes a comparative case study approach, examining five real-world examples of algorithmic decision-making to identify ethical failures and leadership responses. The research combines existing leadership and AI ethics theories to propose a practical framework for organizations to address the moral challenges of AI decision-making. It introduces the "ethical delegation" framework based on intentionality, interpretability, moral authorship, and justice. The goal is to provide theoretical insights and actionable strategies for ethical leadership in various resource environments, bridging the gap between technological advancement and moral responsibility.

## Literature Review

### *Leadership in the Age of Technological Mediation*

Integrating artificial intelligence into organizational decision-making prompts a reevaluation of traditional leadership models that focus on human agency. Theories like transformational leadership (Bass, 1985), servant leadership (Greenleaf, 1977), and ethical leadership (Brown & Treviño, 2006) typically view leaders as the primary moral and decision-making authorities. However, new perspectives indicate that algorithmic systems are emerging as influential autonomous agents, challenging traditional control in organizations (Subrahmanyam, 2025; Syed et al., 2024; Tabata et al., 2025). Subrahmanyam (2025) highlights a shift in leadership towards a data-centric and agile approach instead of individual influence. Tabata et al. (2025) provide a framework showing how generative AI changes decision-making processes and the ethical foundations of leadership. Syed et al. (2024) discuss how AI systems mediate decision-making roles historically held by human leaders.

The ethical dimensions of leadership are being reshaped by digital transformation. While the disruptive effects of AI have been acknowledged (Brynjolfsson & McAfee, 2017), leadership studies are only now starting to explore the ethical implications of integrating intelligent systems. Nyamubarwa's (2025) Ethical Leadership Ecosystem Model offers insights into how organizations can uphold integrity and trust as decision-making becomes shared with algorithms. Additionally, Chen and Ryoo (2025) suggest that fractal AI techniques can improve public health policies, emphasizing that ethical considerations must adapt to technological progress. Chandra (2025) analyzes how AI-driven insights change organizational structures and encourage collaboration beyond traditional leadership roles. Furthermore, Abositta et al. (2024) present evidence that transformational leadership enhances the effectiveness of decision-making in the context of AI, highlighting the need to incorporate the complexities introduced by AI into established leadership theories.

The scholarship indicates a consensus that AI is changing the nature of leadership rather than just serving as a tool for tasks. As organizations adopt AI for decision-making, leadership models will likely evolve into hybrid frameworks that combine human intuition with data-driven precision. This shift will require a rethink of moral guidance and decision-making processes to maintain ethical oversight and effectiveness in an increasingly digital environment.

## *Delegated Judgment and the Ethics of Responsibility*

Delegated judgment has its roots in the ethics of delegation in public administration (Waldo, 1984). As artificial intelligence advances, ethical concerns are shifting from human to algorithmic agency. The concept of the "moral crumple zone" (Elish, 2019) highlights how algorithmic systems complicate accountability in leadership. Leaders now face challenges in moral responsibility when decisions are made through opaque technology, an issue that traditional leadership theories, like the model of distributed morality by Floridi and Sanders (2004), have not adequately addressed.

Recent studies in algorithmic ethics reveal that people generally prefer human moral discretion over making important ethical decisions with algorithms. Jauernig et al. (2022) found that individuals are hesitant to trust algorithms with crucial moral choices, emphasizing the importance of human judgment. Similarly, Villegas-Galaviz and Martin (2023) highlight how the lack of transparency in AI systems can create confusion about who is accountable for ethical decisions, complicating the concept of moral agency and leader responsibility. Kumar et al. (2024) stress the need for increased transparency in AI decision-making, advocating for interpretable models to enhance ethical accountability. This aligns with Vaassen's (2022) critique of the current opaqueness of AI systems and its effects on personal autonomy and ethical responsibility. Overall, these studies underscore the urgent need for frameworks that balance the advantages of algorithmic decision-making with the necessary moral oversight from human leaders. McGuire and Cremer (2022) critically examine the interaction between human decision-making and algorithmic influence, finding that people still prefer human moral judgment over AI despite technological advancements. This skepticism is further supported by Tigard (2020), who argues that ethical decision-making must recognize human and machine contributions, challenging the idea of a "techno-responsibility gap." Peters (2022) adds that the lack of transparency in many algorithms prevents complete alignment with human decisions, highlighting the ongoing importance of human moral judgment even as technology evolves.

The concept of distributed morality provides a valuable foundation for reevaluating ethical leadership in the AI era, but there are notable gaps in its application to leadership contexts. Current literature highlights the need for a comprehensive model that balances the advantages and drawbacks of delegated judgment. This model should emphasize transparency, interpretability, and human oversight to address the ethical risks of algorithmic decision-making, ensuring that moral agency in leadership is preserved.

## AI, Objectivity, and the Illusion of Neutrality

The illusion of objectivity in AI ethics highlights the misconception that algorithmic outputs are neutral. These outputs often reflect biased training data and design choices, undermining their supposed impartiality (Azeem et al., 2023). This bias can mislead leaders and lead to "ethical outsourcing," where responsibility shifts from humans to algorithms assumed to be neutral. As a result, leaders may neglect their moral agency, relying on algorithmic decisions without questioning the underlying biases. Moreover, the idea of algorithmic authority complicates traditional views of leadership legitimacy. According to Beer (2009) and Introna (2015), algorithmic credibility can undermine established models of authority that focus on charisma or legal-rational foundations. Power dynamics may shift toward technical expertise and automated outputs, replacing human judgment with an illusion of rationality. This shift diminishes the moral responsibilities of leaders, prioritizing algorithmic objectivity over ethical accountability. Therefore, decision-makers must utilize transparent, interpretable AI models to critically evaluate technical outputs and maintain ethical oversight as leadership increasingly interacts with algorithms (Azeem al., 2023; Floridi, 2019).

The perception of objectivity in AI ethics hides the biases in algorithmic systems and allows leaders to shift accountability onto these supposedly neutral technologies. Additionally, the rise of algorithmic authority undermines traditional leadership by valuing technical expertise over human ethical considerations. To tackle these challenges, we must promote transparency and interpretability in AI decision-making to uphold ethical leadership in a digital world (Azeem et al., 2023; Floridi, 2019).

## Emerging Debates on AI-Governed Leadership

Recent research is starting to examine the relationship between leadership and Artificial Intelligence (AI), moving away from traditional leadership models toward more integrated approaches. Daugherty and Wilson (2018) discuss "collaborative intelligence," emphasizing that AI can enhance human leadership by improving operational efficiency. However, their focus is mainly on operational aspects, overlooking deeper moral and philosophical issues. In contrast, Joshi (2025) reviews current research on AI in leadership, highlighting the combination of human judgment and machine intelligence for better strategic decision-making. However, like Daugherty and Wilson, this analysis does not fully address the normative questions related to moral authority.

Aziz et al. (2024) systematically review AI-powered leadership, highlighting key themes and ethical challenges related to decision-making with opaque algorithms. They stress the importance of balancing technical benefits with moral responsibility in leadership research. In addition, Pavitra et al. (2024) and Al-Hinaai et al. (2024) explore the evolution of digital leadership. Pavitra et al. (2024) discuss the changes in leadership roles due to digital integration, while Al-Hinaai et al. (2024) assess AI maturity in higher education. Both studies emphasize the need for frameworks that enhance operational outcomes while maintaining ethical standards and accountability in AI-augmented leadership.

Emerging research highlights AI's potential to improve decision-making through collaboration and enhanced strategy, yet there is a notable gap in the ethical delegation of authority. Leaders require a comprehensive framework to manage authority, responsibility, and trust, ensuring that AI's operational advantages align with modern ethical leadership standards. This gap presents an opportunity for research to create ethical guidelines that balance technological progress with strong moral oversight.

## Ethical Delegation Framework: A Normative Model for AI Leadership

"Ethical delegation" involves intentionally transferring decision-making tasks to AI systems, ensuring human leaders retain moral responsibility and oversight. This process requires active judgment, a commitment to human dignity, and engagement with the design and impact of AI decisions. The "ethical delegation" framework in this study provides leaders with guidelines for responsibly assigning decision-making authority to AI while ensuring ethical oversight and moral accountability.

## Core Components of the Framework

To implement the idea of "ethical delegation", the paper proposes these four principles:

1. Delegation should be a conscious moral choice, not just a quick fix.
2. Accountability starts when actions can be understood.
3. Humans must retain moral authorship, even in shared decision-making.
4. Leadership should prioritize justice over efficiency.

These principles provide a framework for assessing and directing leadership behavior in AI-augmented decision-making environments.

### Operationalizing the Framework

Three interrelated theoretical strands inform "ethical delegation":

*Leadership Ethics:* Leaders are responsible for ethical conduct, not just operations (Brown & Treviño, 2006; Greenleaf, 1977). Delegation should reflect principles of integrity, justice, and care.

*Distributed Morality:* According to Floridi and Sanders (2004), decision-making in socio-technical systems involves both human and algorithmic agents, complicating traditional accountability but not removing it.

*Algorithmic Governance and AI Ethics:* There are significant issues of opacity, bias, and ethical ambiguity in AI systems (Elish, 2019; Kumar et al., 2024; Tigard, 2020). These issues require human oversight and clear guidelines for moral responsibility.

These viewpoints indicate that AI leaders must advance from merely adopting technology to fully integrating ethics. The model in Figure 1 guided the thematic coding and comparative analysis of five AI-driven decision-making cases, highlighting leadership responsibility, moral clarity, and contextual outcomes.

**Figure 1**

*Analytical Framework for Assessing Ethical Delegation*

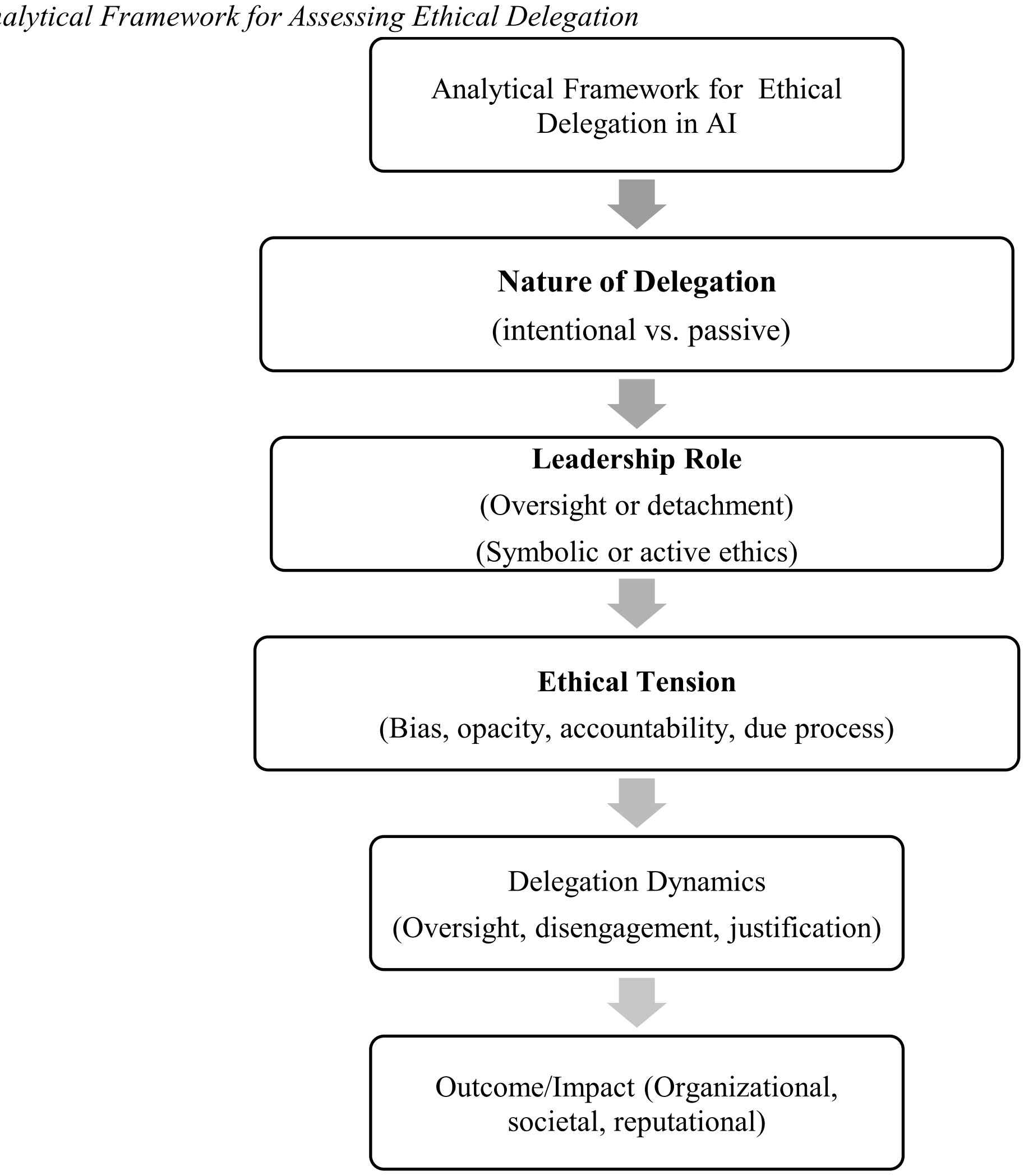

## Positioning Ethical Delegation within Leadership Theories

"Ethical delegation" enhances classical leadership theories—specifically transformational, servant, and ethical leadership—by addressing the new challenges posed by algorithmic systems.

*Extension of Transformational Leadership:* Transformational leadership focuses on vision, influence, and moral purpose (Bass, 1985). "Ethical delegation" builds on this by highlighting the need for moral influence to extend beyond followers to encompass non-human agents, such as algorithms. This demands that leaders proactively consider the ethical implications of the technologies they use to achieve their vision.

*Complement to Servant Leadership:* Servant leadership (Greenleaf, 1977) focuses on the well-being of followers and the community. “Ethical delegation” reinforces this idea by ensuring that AI integration respects stakeholder dignity and avoids harming marginalized groups. It promotes a service-oriented approach through systems stewardship, highlighting the importance of moral care at the infrastructural level.

*Enhancement of Ethical Leadership:* According to Brown and Treviño (2006), ethical leadership emphasizes fairness, integrity, and ethical behavior in relationships. “Ethical delegation” furthers this concept by introducing a technological aspect to moral responsibility. Leaders are now required to address ethical obligations in systems where AI plays a role in decision-making. Unlike traditional models, “ethical delegation” insists that moral responsibility should remain clear and not be hidden by technological complexities or the sharing of decision-making power.

Table 1 summarizes how the “ethical delegation” framework enhances traditional leadership ethics models to meet the new challenges posed by AI-mediated decision-making environments.

**Table 1**
*Comparative Positioning of “Ethical Delegation”*

| Leadership Theory | Core Focus | Ethical Delegation Extension |
|---|---|---|
| Transformational Leadership (Bass, 1985) | Inspires vision and drives moral and strategic change through personal influence. | Adds system-level ethical foresight to vision-setting; requires accountability for algorithmic tools. |
| Servant Leadership (Greenleaf, 1977) | Emphasizes humility, stakeholder care, and community responsibility. | Reframes care ethics around digital systems; leaders must prevent harm through AI governance. |
| Ethical Leadership (Brown & Treviño, 2006) | Models fairness, integrity, and transparency in leader-follower relationships. | Extends accountability into socio-technical systems; moral guidance must apply to algorithmic agents. |
| Ethical Delegation (The Study) | Centers intentional, transparent, and responsible assignment of AI authority. | Synthesizes the above but uniquely addresses leadership in human-AI decision ecosystems. |

## Cases of Delegated Decision-Making in the Era of AI

Artificial intelligence in decision-making poses challenges for leaders across sectors. While AI offers speed, scalability, and objectivity, it can jeopardize traditional leadership by shifting judgment to opaque systems. This chapter examines five case studies that highlight ethical issues in delegating judgment to AI, affecting leadership authority, accountability, and legitimacy. An analytical framework reveals key insights into the ethics of AI delegation.

## Case 1: UK A-Level Grading Algorithm Scandal (2020)

The UK government canceled face-to-face A-level exams due to the COVID-19 pandemic. The Department for Education and Ofqual developed an algorithm to estimate student grades, but the results sparked public outrage (Figure 2). The algorithm was criticized for disproportionately downgrading grades for students from state schools while favoring those from private schools, which disadvantaged lower socio-economic students. This inequity arose from the algorithm's handling of small group sizes, resulting in private schools seeing a greater increase in the number of students achieving top grades (Cowburn, 2025).

**Figure 2**

*Ofqual A-Level Case*

Explainer

Ofqual's A-level algorithm: why did it fail to make the grade?

There is a lot we can learn from the algebraic symbols used to determine results in England

A university vice-chancellor's diary of A-level chaos

Source: The UK Guardian, 2020

*Consequences:* The government initially supported the algorithm for grading to address grade inflation. However, public protests and legal challenges forced a return to teacher-assessed grades, damaging trust in both educational leadership and algorithmic governance (Timmins, 2021)

*Nature of the Delegation:* A statistical AI model was responsible for high-stakes academic assessments based on historical data, school rankings, and teacher predictions. While this delegation was clear, it lacked transparency. The Education Secretary of State at the time, Gavin Williamson, appeared to place blame on Ofqual and emphasized that he was not aware of the scale of the problem.

*Leadership Involvement:* Senior officials supported the algorithm before its release, highlighting consistency in the system, but lacked ethical oversight.

*Ethical Tensions:* Ofqual stated no grading bias (Ofqual, 2020). However, the algorithm favored students from more privileged backgrounds and unfairly penalized those from disadvantaged backgrounds, perpetuating structural inequities and overlooking individual circumstances (Smith, 2020).

The case exemplifies a lack of ethical foresight. Leadership appears to have conflated statistical consistency with fairness, disregarding distributive justice and moral responsibility.

## *Case 2: Amazon's AI Hiring Tool (2014–2017)*

Amazon developed an AI tool to automate résumé screening in recruitment, assigning candidates scores from one to five stars, similar to product ratings. However, by 2015, it was discovered that the system was biased against female candidates for software development and other technical roles. This bias arose because the AI was trained on resumes submitted over a 10-year period, which predominantly came from men, reflecting the male-dominated tech industry. As a result, the system concluded that male candidates were preferable (Dastin, 2018) (see Figure 3).

**Figure 3**
*Amazon AI Recruiting Case*

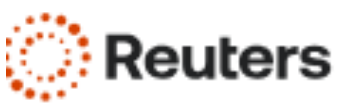

World Business Markets Sustainability Legal Breakingviews Technology Inv

Insight - Amazon scraps secret AI recruiting tool that showed bias against women

By Jeffrey Dastin

October 11, 2018 2:50 AM GMT+2 · Updated 7 years ago

Aa

Source: Reuters, 2018

*Consequences:* The system was shut down due to a lack of accountability and transparency, which raised concerns about AI bias.

*Nature of the Delegation:* The résumé evaluation was assigned to a machine learning model based on past hiring data.

*Leadership Involvement:* HR and technical leadership tested the system internally before quietly decommissioning it.

*Ethical Tensions:* The model showed gender bias by consistently rating female applicants lower because of historical data favoring male hires.

Amazon's leadership failed to recognize and address the ethical issues related to historical bias. They delegated tasks without adequate testing or ethical review.

## Case 3: The COMPAS Algorithm in U.S. Criminal Justice

The COMPAS algorithm was created to assess recidivism risk, influencing bail and sentencing in U.S. courts. A ProPublica analysis by Jeff et al. (2016) revealed that black defendants were more likely than white defendants to be wrongly assessed as high risk for recidivism, while white defendants were more often misclassified as low risk.

**Figure 4**
*Compas Tool Case*

Newsletters The Atlantic

A Popular Algorithm Is No Better at Predicting Crimes Than Random People

The COMPAS tool is widely used to assess a defendant's risk of committing more crimes, but a new study puts its usefulness into perspective.

Source: The Atlantic, 2018

*Consequences:* COMPAS has been heavily criticized for potential racial discrimination in its predictions, as reported in Figure 4. Studies show its accuracy can be as low as 68 percent (Beriain, 2018; Grgić-Hlača et al., 2018). Despite these concerns, the system is still operating, and discussions about algorithmic fairness and judicial responsibility continue.

*Nature of the Delegation:* Judicial discretion was partially assigned to algorithmic risk assessments.

*Leadership Involvement:* Courts and policymakers approved the tool without fully understanding or scrutinizing how it works.

*Ethical Tensions:* Investigations found racial bias, showing that Black defendants were often assigned higher risk scores, raising concerns about due process, fairness, and systemic discrimination.

The situation underscores the potential pitfalls associated with an overreliance on algorithmic tools within the justice system. It illustrates how leaders may inadvertently transfer their moral responsibility by placing undue faith in the perceived objectivity of these tools.

## Case 4: Google and Timnit Gebru's Firing (2020)

Figure 5 showed how Timnit Gebru, a prominent AI ethics researcher at Google, was fired for raising concerns about the social risks of large language models. She co-authored a significant paper showing that facial recognition technology is less accurate for women and people of color, potentially leading to discrimination. Gebru also co-founded the Black in AI group and advocates for diversity in tech. An investigation by the MIT Technology Review revealed that her dismissal was tied to a conflict over another paper she coauthored. In an internal email, Jeff Dean, head of Google AI, stated that the paper "did not meet our bar for publication."

**Figure 5**

*Google Researcher's Case*

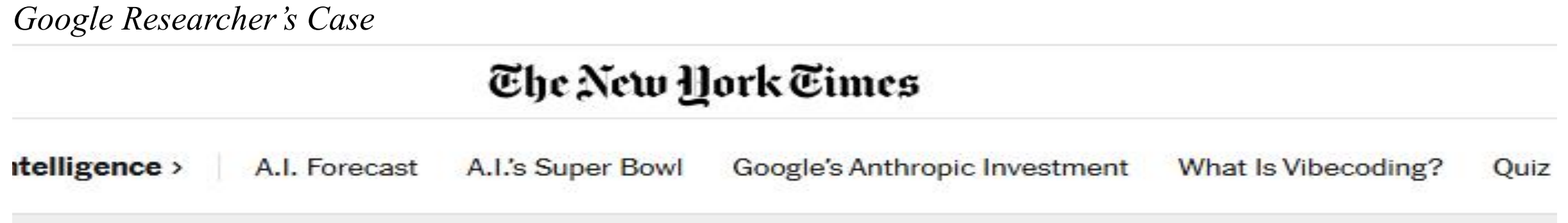

**Google Researcher Says She Was Fired Over Paper Highlighting Bias in A.I.**

Timnit Gebru, one of the few Black women in her field, had voiced exasperation over the company's response to efforts to increase minority hiring.

The New York Times, 2020

*Consequences:* Gebru's dismissal sparked global outrage, prompting resignations and calls for accountability in AI ethics. Many AI ethics leaders argue she was pushed out for revealing uncomfortable truths about the company's research and finances. Over 1,400 Google employees and 1,900 supporters signed a protest letter (Hao, 2020).

*Nature of the Delegation:* The case focuses on delegating ethical inquiry and governance within a corporate structure without involving a specific AI tool.

*Leadership Involvement:* Senior executives overrode internal ethical concerns in favor of reputational and strategic interests.

*Ethical Tensions:* The suppression of ethical dissent and poor transparency have raised global concerns about tech governance and the protection of whistleblowers.
Leadership did not create a space for ethical discussions. Assigning ethics responsibilities to internal teams was merely symbolic and lacked real power.

### Case 5: Dutch Child Welfare Fraud Algorithm (2013–2020)

The Dutch tax authority implemented the SyRI (System Risk Indication) algorithm to detect welfare fraud. When a government agency suspects fraud in a particular neighborhood, it can use SyRI to identify which citizens require further investigation.

**Figure 6**
*Dutch Childcare Case*

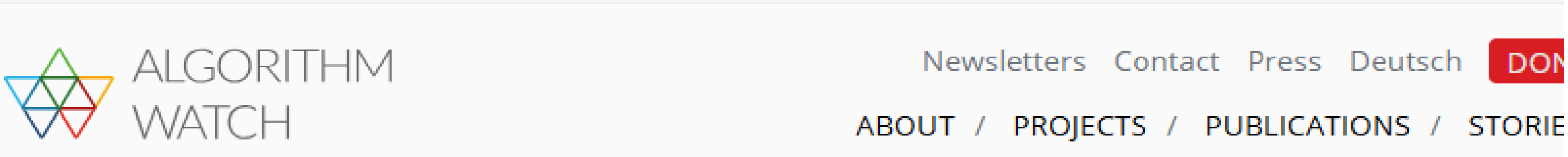

**How Dutch activists got an invasive fraud detection algorithm banned**

**The Dutch government has been using SyRI, a secret algorithm, to detect possible social welfare fraud. Civil rights activists have taken the matter to court and managed to get public organizations to think about less repressive alternatives.**

Source: AlgorithmWatch (2020)

*Consequences:* The program was discontinued due to public outrage and mass resignations from the Dutch tax authority, significantly damaging institutional credibility (Figure 6). It disproportionately affected minority and low-income families. Research by the Dutch news media, De Volkskrant, in 2019, found that algorithmic investigations did not uncover new fraud cases (Huisman, 2019). The research also revealed false positives and a lack of transparency (AlgorithmWatch, 2020). In a letter to the Dutch court on September 26, 2019, Philip Alston, the United Nations Special Rapporteur on extreme poverty and human rights, stated that entire neighborhoods were unfairly scrutinized for fraud while wealthier areas faced no oversight (AlgorithmWatch, 2020). A parliamentary inquiry concluded that the program violated fundamental principles of the rule of law.

*Nature of the Delegation:* Fraud detection and risk scoring were automated with little human oversight.

*Leadership Involvement:* Senior officials ignored warning signs and ethical concerns, pushing forward with the program without a proper investigation. An internal probe revealed that in November 2019, a former Tax and Customs Administration employee sent an urgent letter to the House of Representatives. This employee, who processed objections for the Benefits department from 2014 to 2016, highlighted the unfair treatment of parents and a lack of legal basis for the activities. He reported these concerns to his supervisor multiple times, but no action was taken.

*Ethical Tensions:* Discrimination, lack of due process, and unfair penalties reveal significant moral failures in design and governance.
Leadership's passive acceptance of algorithmic authority resulted in significant social harm and political consequences.

## Comparative Analysis of Cases

Table 2 organizes the five case studies across key analytical dimensions of AI use, leadership role, ethical failure, and delegation dynamics. These five cases highlight a significant trend: leaders who rely on AI systems without proper ethical oversight can face serious consequences. Blind trust in algorithms, insufficient questioning of design assumptions, and neglecting ethical issues undermine moral leadership. Leaders must grasp the tools and maintain moral authority to lead effectively in an AI-driven environment, ensuring that delegating tasks does not mean neglecting responsibility. These patterns reveal a deeper issue—moral drift—where leadership loses ethical clarity through excessive delegation of technical tasks. The following section examines the impact on leadership training and institutional design.

**Table 2**
*Case Comparison Matrix*

| Case | AI System Type | Leadership Role | Ethical Dilemma | Delegation Dynamics | Outcome/Impact |
|---|---|---|---|---|---|
| Amazon AI Hiring Tool | Resume-screening NLP | HR and tech leadership designed and deployed the system | Gender bias in hiring decisions | Delegation to AI for pre-interview filtering without adequate bias auditing | System discontinued quietly; limited public accountability. |
| COMPAS (US Criminal Justice) | Risk prediction algorithm | Judges and justice administrators used tools in sentencing | Racial bias, unjust outcomes | Human judges deferred to algorithmic risk scores | System remains in use; intense public scrutiny and ethical criticism |
| UK A-Level Grading Algorithm | Predictive grading algorithm | Government education officials approved and implemented it | Undermining merit, penalizing disadvantaged students | Delegation of national grading authority to opaque AI model | Public outcry, government reversed decision and apologized |
| Google-Timnit Gebru Conflict | Ethical AI research and internal governance | Tech leadership overrode internal ethics voices | Ethical suppression, lack of transparency | Delegation of ethical oversight subordinated to corporate image | Global backlash raised awareness of ethics suppression in tech |
| Dutch Welfare Fraud Scandal | Fraud detection algorithm | Tax authorities and ministers failed to intervene | Discriminatory profiling of families | Over-dependence on AI flags with minimal human verification | Mass resignations; state apology; long-term reputational damage |

# Method

## Data Sources & Case Selection

We analyzed four significant AI deployments: the UK A-Level grading system, Amazon's recruiting tool, COMPAS recidivism scores, and the SyRI welfare-fraud system. We employed purposive sampling to capture variations in fields such as education, employment, criminal justice, and welfare, with a focus on the ethical implications of each case. For each, we collected: Official documents (e.g., government reports, technical specifications; N=24); and Media and trade-press articles (from outlets such as The Guardian, Reuters, New York Times; N=37)

Cases were chosen based on three criteria: (a) clear leadership involvement, (b) public access to materials, and (c) substantial real-world impact.

## Coding Framework & Procedures

We conducted a two-phase grounded-theory analysis on all collected texts using NVivo 12.:
*Phase I:* Two coders analyzed a 20% random subset of documents, leading to 46 initial codes.
*Phase II:* We developed a codebook through iterative calibration meetings, resulting in 32 defined codes (e.g., *delegation intent, stakeholder explainability, decision ownership, equity assessment*), each with clear criteria and example quotes. The codes were organized into four main themes:

1) Intentionality: Measures how often leaders intentionally delegate decisions to AI by monitoring formal policies, approval processes, escalation protocols, and data quality checks.
2) Interpretability: Captures efforts to make algorithmic logic clear and understandable for stakeholders, including explainability materials, clear documentation, transparency measures, and feedback channels.
3) Moral Authorship: Evaluates how and if individuals take responsibility for outcomes affected by AI, including decision ownership, ethical guidelines, error reporting, and human override systems.
4) Justice: Implement checks for unfair or biased outcomes, including equity assessments, bias detection, fairness metrics, privacy safeguards, and stakeholder engagement.

Table 3 presents a 32-item codebook divided into three columns: Code, Theme, and Description.

**Table 3**
*The 32-Item Codebook of Thematic Content Analysis (Developed by this Research)*

| Codes | | Themes | Descriptions |
|---|---|---|---|
| 1 | accountability_frameworks | | Structures (e.g., committees, policies) that assign responsibility for AI outcomes. |
| 2 | audit_trail | | Comprehensive logging of all decisions, data versions, and model iterations for retrospective review. |
| 3 | contextual_adaptation | | Customizing AI tools to fit specific organizational or cultural settings. |
| 4 | data_quality | | Assessments of the accuracy, completeness, and relevance of input data. |
| 5 | delegation_intent | | Explicit decision by leaders to involve AI rather than defer unconsciously. |
| 6 | escalation_protocols | | Defined steps for raising critical issues to higher authority. |
| 7 | governance_mechanisms | | Organizational processes for oversight and control of AI projects. |
| 8 | interdisciplinary_collaboration | | Cooperation between technical, legal, and ethical experts. |
| 9 | model_validation | | Procedures to test model performance against benchmarks or real-world outcomes. |
| 10 | performance_metrics | | Quantitative indicators used to monitor AI accuracy, precision, recall, etc. |
| 11 | resource_allocation | | Decisions about funding, staffing, and time dedicated to AI governance. |

| 12 | risk_assessment | Intentionality | Identification and analysis of potential harms from AI deployment. |
|---|---|---|---|
| 13 | technical_infrastructure | | Availability and maintenance of hardware/software enabling AI operation. |
| 14 | training_procedures | | Methods for preparing coders and analysts to apply the codebook reliably. |
| | | | |
| 15 | stakeholder_explainability | Interpretability | Efforts to make AI logic understandable to affected parties. |
| 16 | stakeholder_feedback | | Channels for end-users and affected individuals to comment on AI decisions. |
| 17 | transparency_measures | | Documentation and disclosure practices about model design and use. |
| 18 | documentation_clarity | | Quality and readability of written materials describing methods and tools. |
| 19 | decision_latency | Moral Authorship | Time between AI recommendation and final human decision. |
| 20 | decision_ownership | | Attribution of the final decision outcome to a specific human actor. |
| 21 | ethical_guidelines | | Formal principles or codes guiding responsible AI use. |
| 22 | error_reporting | | Systems for logging, investigating, and responding to AI errors or failures. |
| 23 | human_override | | Mechanisms enabling humans to review and reverse AI recommendations. |
| 24 | legal_compliance | | Alignment of AI use with relevant laws and regulations. |
| 25 | moral_authorship | | Leader's explicit claim of moral responsibility for AI-informed decisions. |
| 26 | equity_assessment | Justice | Evaluation of whether AI-driven decisions produce fair outcomes across groups. |
| 27 | bias_detection | | Processes to identify and flag systematic errors or prejudices in the model. |
| 28 | fairness_metrics | | Statistical measures (e.g., demographic parity, equalized odds) used to quantify fairness. |
| 29 | impact_monitoring | | Ongoing tracking of real-world consequences after deployment. |
| 30 | privacy_safeguards | | Measures to protect sensitive information in datasets and models. |
| 31 | stakeholder_engagement | | Involvement of affected groups in design, evaluation, or governance phases. |
| 32 | user_consent | | Processes ensuring individuals agree to their data being used by AI. |

## Analysis

We conducted a matrix coding query on 110 coded text segments to support our insights. Table 4 shows the absolute counts and relative frequencies for each of the four themes. We then performed:

1) Inter-Coder Reliability Check – mean Cohen's $\kappa$ = .82 across themes, indicating substantial agreement.
2) Chi-Square Test – $\chi^2(3) = 3.84$, $p = .28$, showing no significant variation in theme application across the four cases.

**Table 4**

*Theme Frequencies in Thematic Coding (N = 110)*

| Theme | Count | Percentage (%) |
|---|---|---|
| Intentionality | 28 | 25.5 |
| Interpretability | 23 | 20.9 |
| Moral Authorship | 31 | 28.2 |
| Justice | 28 | 25.5 |

The results demonstrate that all themes are supported by strong empirical evidence, with no theme being overrepresented.

*Summary of Results*

To answer our first research question—*How can leaders responsibly assign judgment to AI systems?* —our analysis shows that effective delegation relies on:

1) Intentionality: Leaders should actively choose to use AI and document that decision, instead of allowing AI to make decisions passively (e.g., UK A-Level grading; Amazon resume screening).
2) Interpretability: Delegation needs models with clear logic to explain to stakeholders, helping to anticipate bias (like COMPAS recidivism scores) and justify decisions (such as SyRI welfare-fraud flags).

Our second question is: *What core principles define 'ethical delegation,' and how do they maintain accountability in AI-mediated decision-making?* —is answered by:

3) Moral Authorship: Leaders must take full responsibility for final decisions, even when based on algorithms input (as seen when the Education Secretary reinstated teacher grades during the A-Level scandal).
4) Justice: Delegation should be adjusted to ensure that it does not reinforce structural inequities, like the negative effects on low-income parents in the Dutch welfare fraud system.

## Discussion

This paper examines five case studies highlighting ethical failures in AI-driven decision-making, emphasizing the need to rethink leadership ethics in the digital age. The findings reveal that leaders often confuse technological efficiency with ethical responsibility, ceding authority to AI without ensuring transparency, accountability, or moral clarity.

Leaders struggle to maintain ethical oversight when using algorithmic tools in various sectors, such as public education, corporate recruitment, criminal justice, technology governance, and welfare administration. These issues often arise from overreliance on perceived objectivity, uncritical acceptance of AI outputs, and a lack of readiness to challenge or limit AI systems.

The concept of "ethical delegation" proposed in this study offers a new perspective on leadership. It advocates for intentional delegation, maintaining moral responsibility, and ongoing oversight to prevent harm and reinforce the proper moral purpose of leadership.

### *Implications for Leadership Training and AI Governance*

*Leadership Education:* Leaders must receive ethics education, including training in digital accountability, understanding algorithmic bias, and making socio-technical decisions. This training should go beyond basic AI literacy to encompass moral reasoning in human-machine interactions.

*Governance Frameworks:* Organizations must incorporate "ethical delegation" principles into their AI governance policies. Before deploying algorithms, this should include guidelines for explainability, redress, stakeholder consultation, and transparent risk assessments.

*Organizational Culture:* "Ethical delegation" requires an organizational culture prioritizing moral responsibility in AI systems. This includes establishing cross-disciplinary ethics boards to oversee AI use, conducting algorithmic impact audits to identify social risks, and employing AI red-teaming to test for bias and failure. Organizations must offer protected channels for employees to report ethical concerns and integrate ethics into leadership training. These practices turn ethics from mere symbols into concrete actions, promoting accountability, transparency, and moral awareness in AI-driven decision-making.

AI does not diminish leadership; it raises its importance. Leaders are now not just decision-makers but also curators of decision-making systems that involve non-human agents.

### Limitations

Some key limitations should be acknowledged:

*Case Scope and Generalizability:* The study focuses on five carefully chosen, high-profile cases primarily from Western contexts. While these cases are detailed and illustrative, they may not fully capture the range of AI ethics challenges, especially in Global South regions or less-publicized industries.

*Public Data Reliance:* All case analyses depend on secondary sources such as news reports, academic literature, and public records. This limits access to internal discussions, ethical reviews, and detailed leadership reasoning that could be better understood through interviews or ethnographic methods.

*Evolving Technological Landscape:* AI technologies and their ethical implications are changing quickly. This study represents a snapshot rather than a comprehensive or future-proof analysis. Future technological and regulatory changes may significantly alter the ethical landscape.

Future research should apply this framework to new sectors like climate governance, military AI, and humanitarian aid, using interviews or ethnographic methods to enhance understanding of leadership reasoning and ethical intent.

## Conclusion

This paper meaningfully contributes to the emerging field of AI and leadership ethics by presenting "ethical delegation" as a normative framework for decision-making in AI-driven environments. It emphasizes that the fundamental challenge of algorithmic leadership lies not solely in technical expertise but in maintaining moral accountability within distributed decision-making systems.

The case studies reveal that relinquishing judgment to machines without preserving moral authorship results in ethical deterioration, reduced legitimacy, and potential institutional failure. Leaders must adapt not by shying away from technology but by embracing it with a critical and ethical mindset.

The future of leadership in the age of AI hinges not on mere technical prowess but on moral clarity. As intelligent systems increasingly influence decisions that impact human lives, leaders must prioritize human values at the core of digital governance. "Ethical delegation" presents a viable path forward that affirms accountability, strengthens trust, and reclaims the essential human nature of leadership. Without a clear ethical framework, intelligent systems could lead to moral drift. Leadership must not just keep up with technology; it must also take charge of guiding it.

## Declarations

## Acknowledgements

Not applicable.

## Disclosure Statement

No potential conflict of interest was reported by the authors.

## Ethics Approval

Not applicable.

## Funding Acknowledgements

Not applicable.

## Citation to this article

## Rights and Permissions

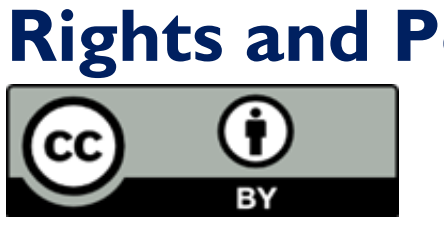